\documentclass{PoS}

\newcommand\bef{\begin{figure}}
\newcommand\eef[1]{\label{fg:#1}\end{figure}}
\newcommand\beq{\begin{equation}}
\newcommand\eeq[1]{\label{#1}\end{equation}}
\newcommand\beqa{\begin{eqnarray}}
\newcommand\eeqa[1]{\label{#1}\end{eqnarray}}
\newcommand\bet{\begin{table}}
\newcommand\eet[1]{\label{tb:#1}\end{table}}

\newcommand\fgn[1]{Figure \ref{fg:#1}}

\title{Absence of bilinear condensate in three-dimensional QED}

\ShortTitle{Absence of bilinear condensate in three-dimensional QED}

\author{\speaker{Nikhil Karthik}\\
        Florida International University\\
        E-mail: \email{nkarthik@fiu.edu}}

         \author{Rajamani Narayanan\\
         Florida International University\\
         E-mail: \email{rajamani.narayanan@fiu.edu}}

\abstract{ 
There are plausibility arguments that QED in three dimensions has
a critical number of flavors of massless two-component fermions,
below which scale invariance is broken by the presence of bilinear
condensate. We present numerical evidences from our lattice
simulations using dynamical overlap as well as Wilson-Dirac fermions for the
absence of bilinear condensate for any even number of flavors of two-component fermions. 
Instead, we find evidences for the scale-invariant nature of three-dimensional QED.
}

\FullConference{34th annual International Symposium on Lattice Field Theory\\
		24-30 July 2016\\
		University of Southampton, UK}

\begin{document}

\section{Introduction}
Parity-invariant QED$_3$ with $2N_f$ flavors of massless two-component
fermions coupled to three-dimensional non-compact Abelian gauge-fields
has been studied in the past as a quantum field theory which can be tuned
to be conformal or to have a mass-gap by changing $N_f$. The question is
the following -- is there a critical number of flavors of two-component
fermions $2N_f$ below which massless QED$_3$ in a finite box of length
$\ell$ generates other low-energy length scales which are independent
of $\ell$ as $\ell\to\infty$? One such low-energy length scale that
is of interest is the bilinear condensate $\Sigma$ which, if non-zero,
governs the following scaling of the low-lying eigenvalues $\lambda_i$
of the massless Dirac operator:
\begin{equation}
\lambda_i = \frac{z_i}{\Sigma} \frac{1}{\ell^3}, 
\end{equation}
where $z_i$ are universal numbers depending only on the symmetries
of the Dirac operator, and can be obtained from a random matrix
model with the same symmetries (refer~\cite{Verbaarschot:1994ip}
for such a model corresponding to QED$_3$). In this talk, based on our
publications~\cite{Karthik:2016ppr,Karthik:2015sgq}, we primarily address
the existence of $\Sigma$ for small $N_f$ ($=1,2,3,4$) by asking if
$\lambda\sim\ell^{-(1+p)}$ with $p=2$.  We summarize the status of the
understanding of the critical $N_f$ before our studies in \fgn{status}
(see~\cite{Karthik:2016ppr} and references therein, for a complete
literature survey). The analytical computations, each with their own
limitations, suggested that the critical $N_f$ lie between 0 and 4. The
previous lattice studies suggested that it could be 1 or 2.

\begin{figure}
\centering
\includegraphics[scale=0.42]{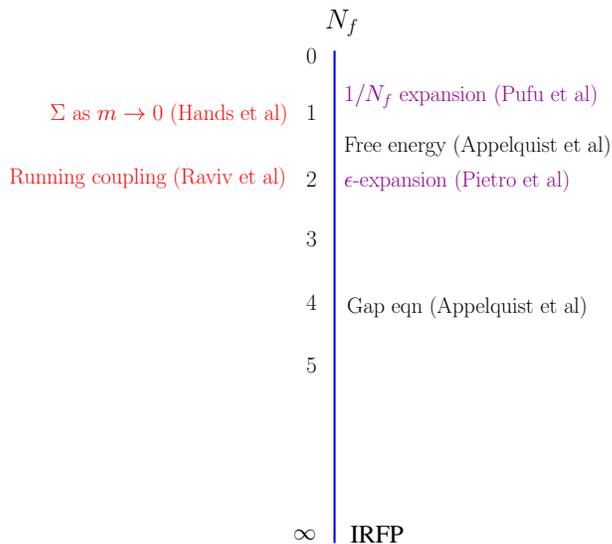}
\caption{Few representative older
calculations~\cite{Appelquist:1986fd,Appelquist:1999hr,DiPietro:2015taa,Chester:2016ref,Hands:2004bh,Raviv:2014xna}
of the critical value of $N_f$ below which bilinear condensate
exists. The large-$N_f$ computation points to an infra-red fixed
point. Various perturbative calculations as well as approximate
solutions to the gap equation have been carried out to investigate
the stability of the infra-red fixed point. These calculations
suggest the critical value might lie anywhere between 0 and 4. The
previous non-perturbative lattice studies of QED$_3$ suggest this
critical value might be 1 or 2.  }

\label{fg:status}
\end{figure}

\section{Lattice details}
We regulated QED$_3$ in a finite box of physical volume $\ell^3$ using  $L^3$ lattices.
The lattice coupling appearing in the gauge action is $\beta=L/\ell$;
the continuum limit at a fixed physical length $\ell$ is taken by
extrapolating to $L\to\infty$. We regulated the two
flavors of massless two-component fermions in a parity-invariant way using the Wilson-Dirac as
well as overlap fermions.  The fermion propagator $G$ for the parity-preserving
Wilson-Dirac fermion is

\beq
G^{-1} = \left[\begin{array}{c c} 0 & X \\ -X^\dagger & 0 \end{array}\right]\quad; \quad  X=C_n+B-m_t.
\eeq{propwil}
$C_n$ is the two-component naive Dirac operator, $B$ is the Wilson
term and $m_t$ is tuned such that the lowest eigenvalue $\lambda_1$
of $i G^{-1}$ is minimum. We further improved it by adding a
Sheikholeslami-Wohlert term and by using HYP smeared links in the
Dirac operator.  The fermion propagator $G$ for the overlap fermion, which has
the full U$(2N_f)$ symmetry even at finite lattice spacing, is given
in terms of a unitary matrix $V=(X^\dagger X)^{-1} X$ as \footnote{The Wilson mass $m_t=1$ in overlap simulations}
\beq
G^{-1} = \left[\begin{array}{c c} 0 & \frac{1-V}{1+V} \\ \frac{1-V}{1+V} & 0 \end{array}\right].
\eeq{propov}
We define the ``eigenvalues of the Dirac operator'' in either case
to be the eigenvalues $\lambda_i$ of $iG^{-1}$ which are real.  We
used standard HMC for generating $\sim 500-1000$ independent gauge
configurations at all the simulation points ($4\le \ell\le 250$).
Using Wilson-Dirac fermions we studied $N_f=1,2,3$ and 4. With the
overlap fermion, we studied $N_f=1$. At each $\ell$, we used multiple
$L^3$ lattices ($12\le L\le 24$) in order to take the continuum
limits.  

\section{Evidence from $\ell$-scaling of the low-lying eigenvalues of Dirac operator}

\begin{figure}
\begin{center}
\includegraphics[scale=0.5]{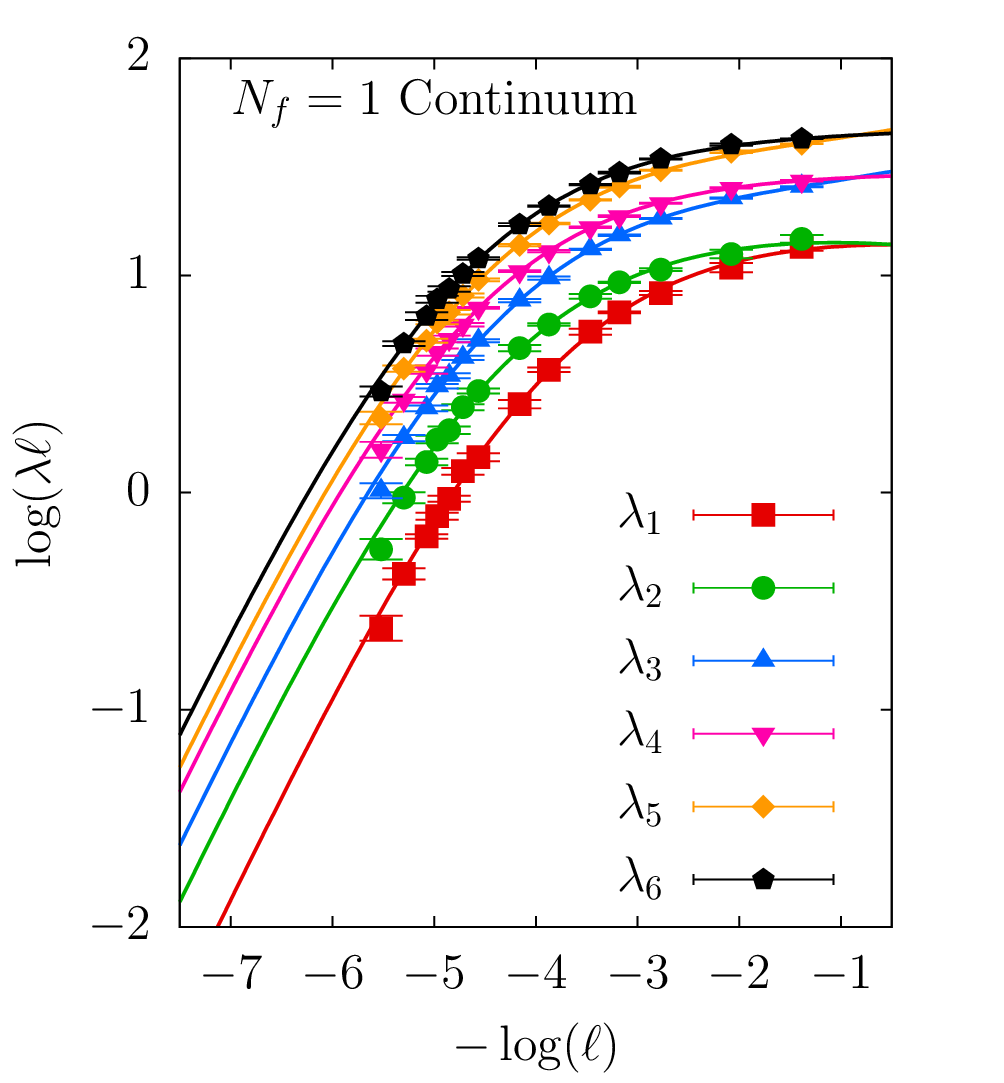}
\includegraphics[scale=0.6]{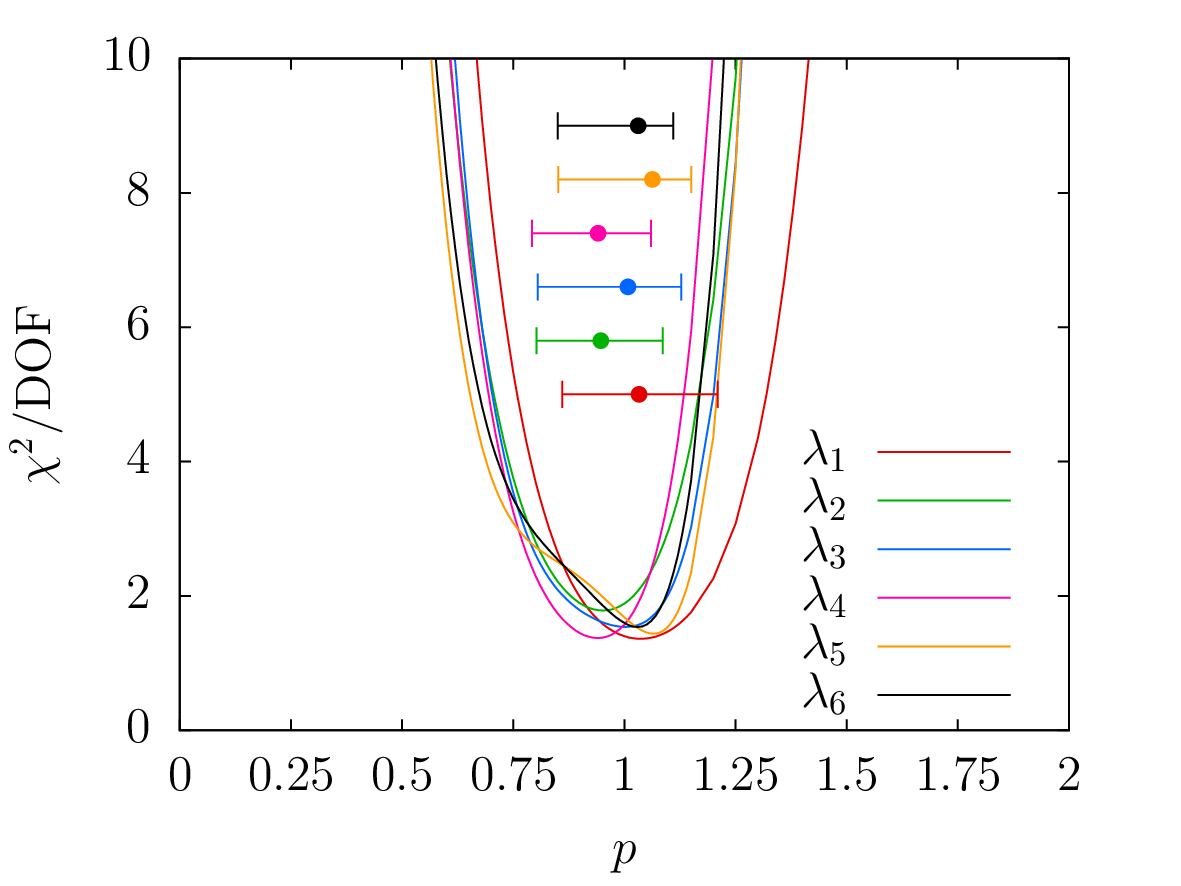}
\end{center}
\caption{On the left panel, the $\ell$ dependence of the six
low-lying, continuum extrapolated, eigenvalues of the overlap
operator is shown. The Pad\'e approximations to their $\ell$ dependence
with $p=1$ are shown as the solid curves. On the right panel, the
likelihood of different values of the exponent $p$, measured
using the $\chi^2/\rm{DOF}$ for the best fit of the Pad\'e approximation 
with various
values of $p$ to the finite $\ell$ data, is shown.
}
\label{fg:ldepnf1}
\end{figure}

In a finite physical box, the spectrum of the Dirac operator is
discrete. Thus, one can talk about the $\ell$-dependence of the
individual low-lying eigenvalues. As we noted in the introduction,
the $i$-th low-lying eigenvalue $\lambda_i$ will scale as $\ell^{-3}$
when there is a condensate $\Sigma$. If $\ell^{-3}$ scaling is not
found, we can conclude that a bilinear condensate is absent and instead
we can obtain the mass anomalous dimension of the
scale-invariant theory; since $\lambda$ has an engineering dimension
of mass, the mass anomalous dimension $\gamma_m$ is $p$ if $\lambda\sim
\ell^{-p-1}$ and $p< 1$.

In the left panel of \fgn{ldepnf1}, we
show the dependence of the continuum extrapolated values of $\lambda_i\ell$
as a function of $1/\ell$ for the six low-lying eigenvalues of the
overlap operator in a log-log plot. At any finite $\ell$ that we
studied, the slope $\frac{d\log(\lambda\ell)}{d\log(1/\ell)}$  is
less than 2, the value that is expected if $\Sigma\ne 0$.  In fact, it is
less than 1. We estimate the exponent of the power-law that would
be seen as $\ell\to\infty$ by describing the $\ell$-dependence of
our data by
\beq
\lambda\ell = \ell^{-p} F(1/\ell),
\eeq{anstz}
with an unknown scaling correction $F$. We approximate $F$ by a
$[1/1]$ Pad\'e approximant.  We find it numerically stable to
write the Pad\'e approximant in terms of $\tanh(1/\ell)$. The best
fits of the above ansatz with $p=1$ to the data are shown by the
solid curves in the left panel of \fgn{ldepnf1}. In the right panel,
we show the $\chi^2/\rm{DOF}$ for such fits to the six low-lying
eigenvalues as a function of the exponent $p$. The value $p=2$ is
clearly ruled out, which implies the absence of a condensate.
Assuming the theory does not generate other length scales as well, we
can estimate the mass anomalous dimension $\gamma_m=p$ of the theory
to be 1.0(2) from the same plot.  Further, we support the correctness
of our result by comparing the $\ell$-dependence of the continuum
extrapolated low-lying eigenvalues of the two different lattice Dirac
operators in \fgn{compare}. A perfect agreement between the
Wilson-Dirac and the overlap formalisms is seen. Due to such an agreement, 
we study the $N_f=2,3,4$ cases using only the Wilson-Dirac fermion.
\begin{figure}
\begin{center}
\includegraphics[scale=0.49]{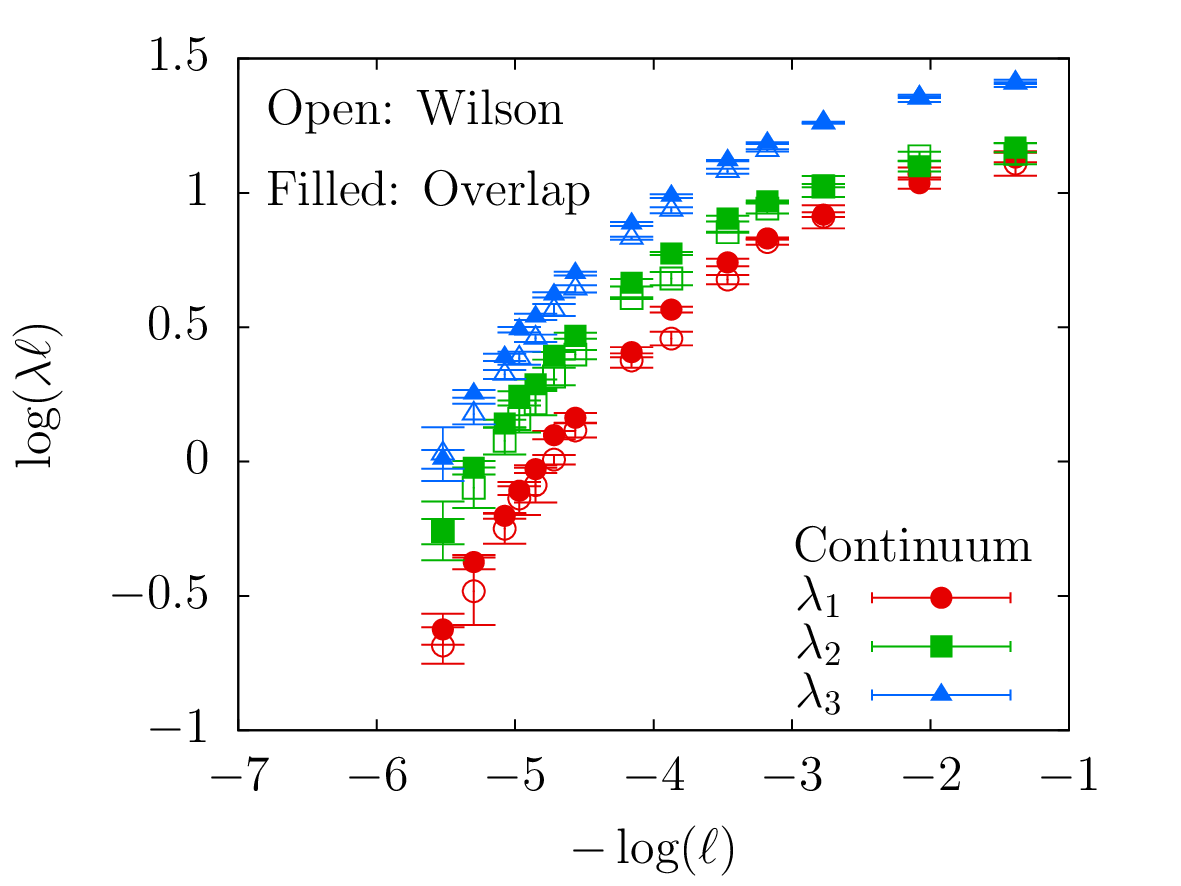}
\end{center}
\caption{The plot compares the $\ell$-dependence of the first three
low-lying eigenvalues, after taking the continuum limit, using
Wilson fermions (open symbols) and overlap fermions (filled symbols)
for the $N_f=1$ case.  }
\label{fg:compare}
\end{figure}

\begin{figure}
\begin{center}
\includegraphics[scale=0.45]{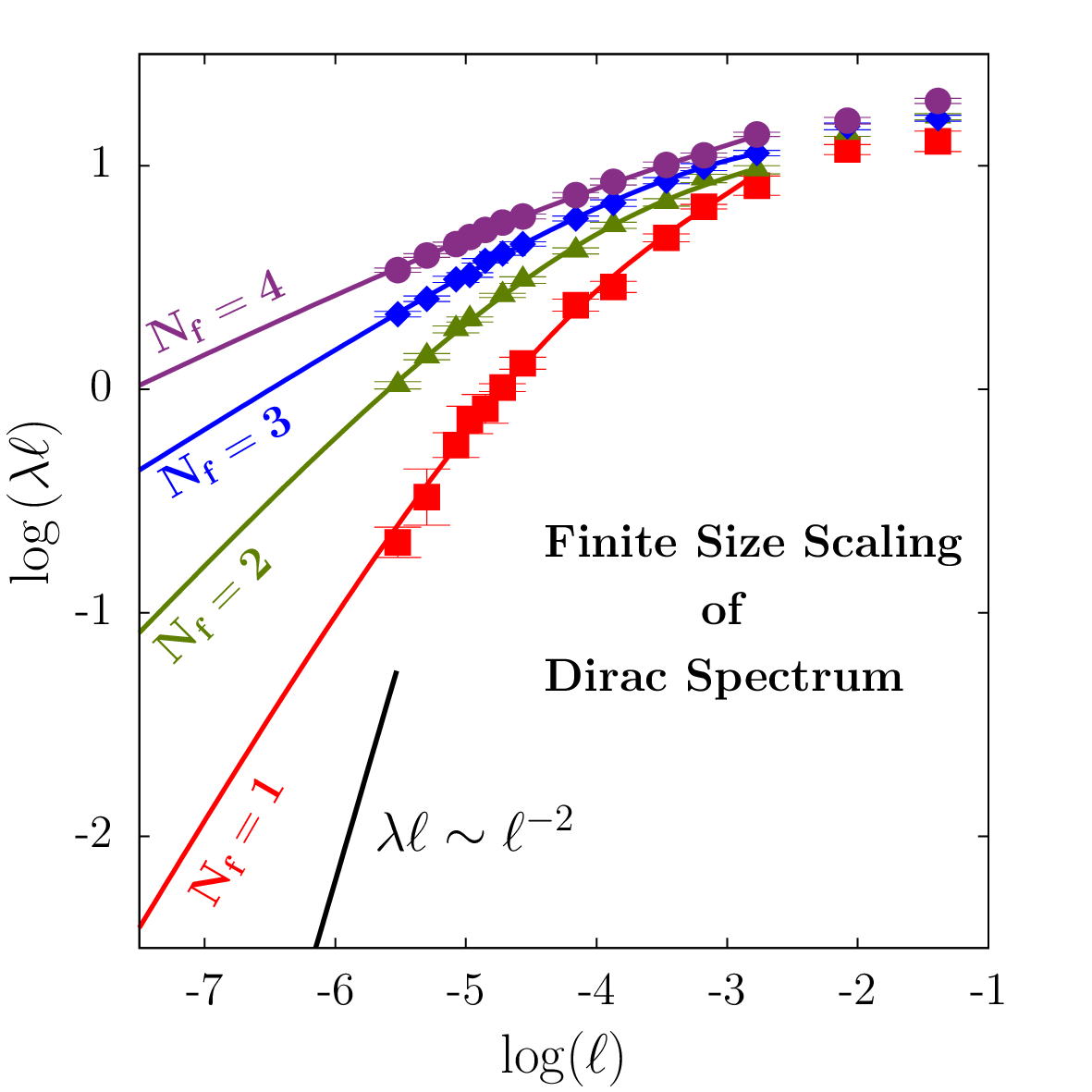}
\end{center}
\caption{The $\ell$-dependence of the smallest eigenvalue of the
Wilson-Dirac operator for $N_f=1,2,3$ and 4.  The expected scaling
when a bilinear condensate is present, $\lambda\ell\sim \ell^{-2}$,
is shown by the black straight line in this log-log plot. The
exponent $p$ for the asymptotic $\ell$-scaling seems to decrease
as $1/N_f$.
}
\label{fg:multi}
\end{figure}

In \fgn{multi}, we show the $\ell$-dependence of the continuum
extrapolated smallest eigenvalue for different number of flavors
$N_f=1,2,3$ and 4. The eigenvalues scale with a smaller exponent
$p$ as $N_f$ increases, consistent with the expectation that if
$N_f=1$ does not have a bilinear condensate, the $N_f=2,3,4$ also
would not. Thus QED$_3$ does not have a bilinear condensate for all
non-zero $N_f$. Again, assuming this means that QED$_3$ is
scale-invariant for all $N_f$, we estimate the mass anomalous
dimension to be $\gamma_m=1.0(2), 0.6(2), 0.37(6)$ and 0.28(6) for
$N_f=1,2,3,4$ respectively. Surprisingly, this agrees with an analytical 
calculation~\cite{Gracey:1993sn} of $\gamma_m$ to $\mathcal{O}(1/N_f^2)$ where no assumption about 
bilinear condensate is made; the analytical values are 
$\gamma_m=1.19, 0.56,0.37$ and 0.28 for $N_f=1,2,3,4$ respectively.

\begin{figure}
\begin{center}
\includegraphics[scale=0.45]{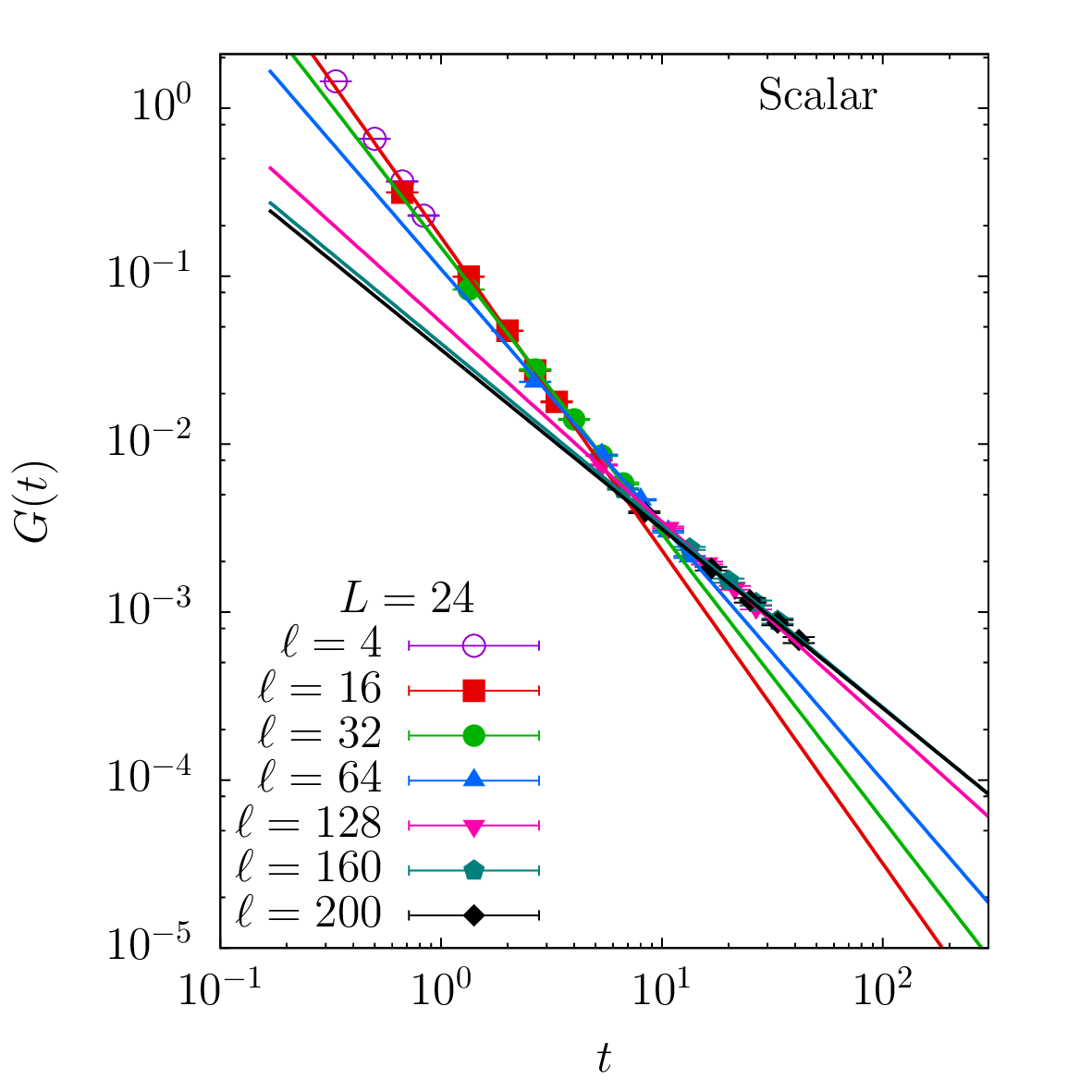}
\includegraphics[scale=0.55]{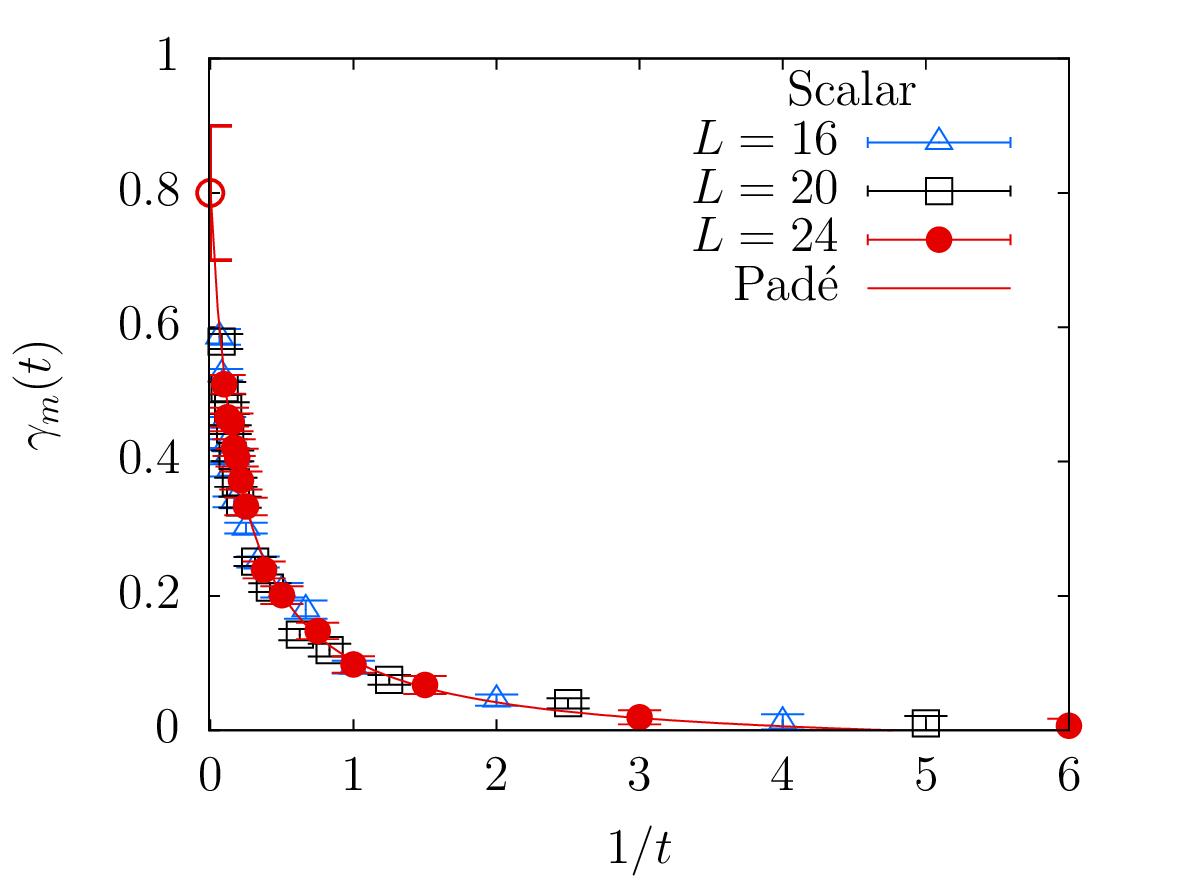}
\end{center}
\caption{
(Left) The zero spatial momentum scalar correlator
$G(t)=\left\langle\Sigma(0)\Sigma(t)\right\rangle$ as a function
of temporal separation $t$. The different lines are tangents to the
correlator, with slope $k(t)$, at various $t$ on the log-log plot.
(Right) The mass anomalous dimension given by $\gamma_m(t)=1-k(t)/2$
is plotted as a function of the scale $t$. 
}
\label{fg:correl}
\end{figure}

The other way to obtain the mass anomalous dimension
is to study the scalar correlator
$G(t)=\left\langle\Sigma(0)\Sigma(t)\right\rangle$ projected to
zero spatial momentum. The correlator is shown as a function of the
temporal separation $t$ in the left panel of \fgn{correl}. The first
thing to notice is the concave-up nature of the correlator. This
indicates the absence of a mass-gap, thereby ruling out the presence
of another length scale in addition to a bilinear condensate. The
slope on the log-log plot, $k(t)=\frac{d\log(G(t))}{d\log(t)}$, is
related to a scale dependent mass anomalous dimension $\gamma_m(t)$
as $\gamma_m(t)=1-k(t)/2$.  This is shown as a function of $1/t$
in the right panel of \fgn{correl}. The mass anomalous dimension
at the IR fixed point to which QED$_3$ with $N_f=1$ flows to, is
$\gamma^*=\lim_{t\to\infty}\gamma_m(t)$.  We estimate by an
extrapolation that $\gamma^*=0.8(1)$. This is consistent with the
estimate 1.0(2) from the eigenvalues described above. The agreement
between two different approaches to $\gamma^*$ serves as a cross-check.

\section{Evidence from Inverse Participation Ratio and number variance}
\begin{figure}
\begin{center}
\includegraphics[scale=0.6]{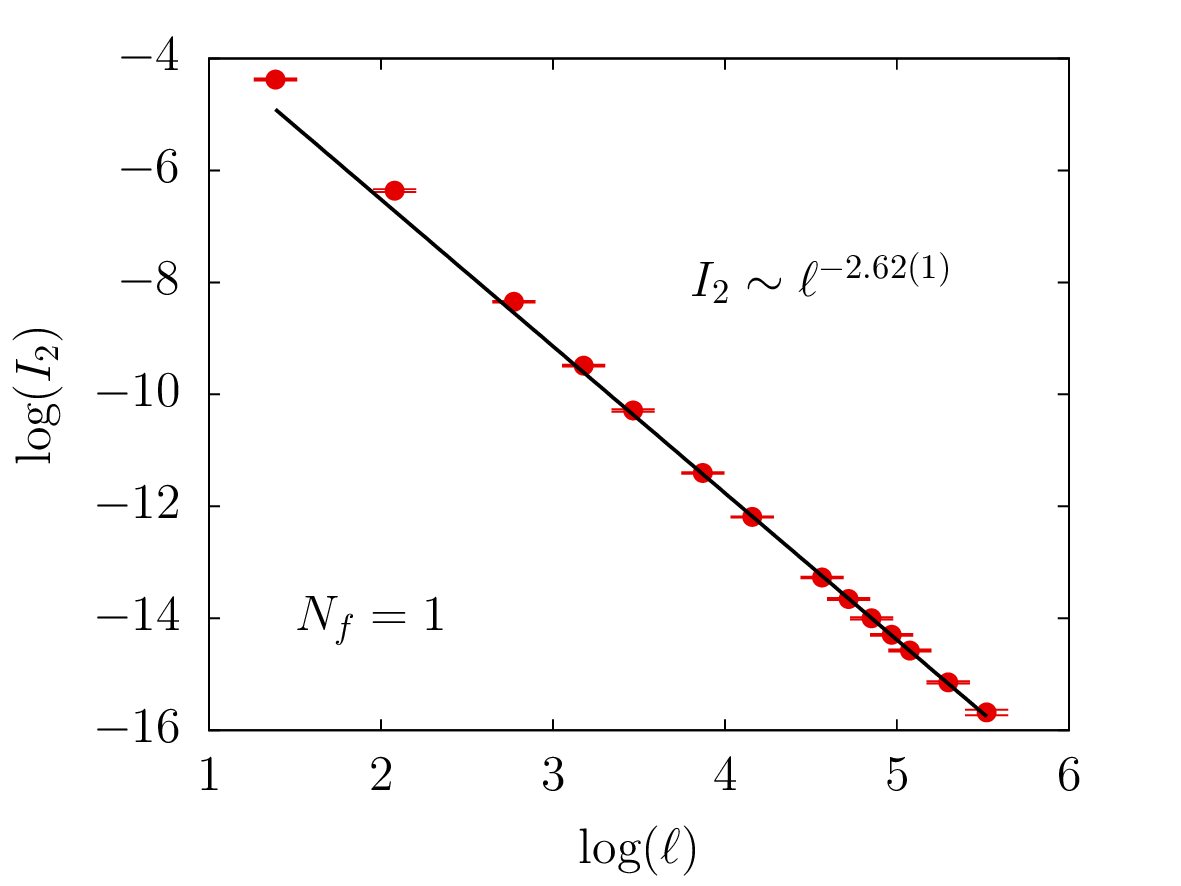}
\includegraphics[scale=0.5]{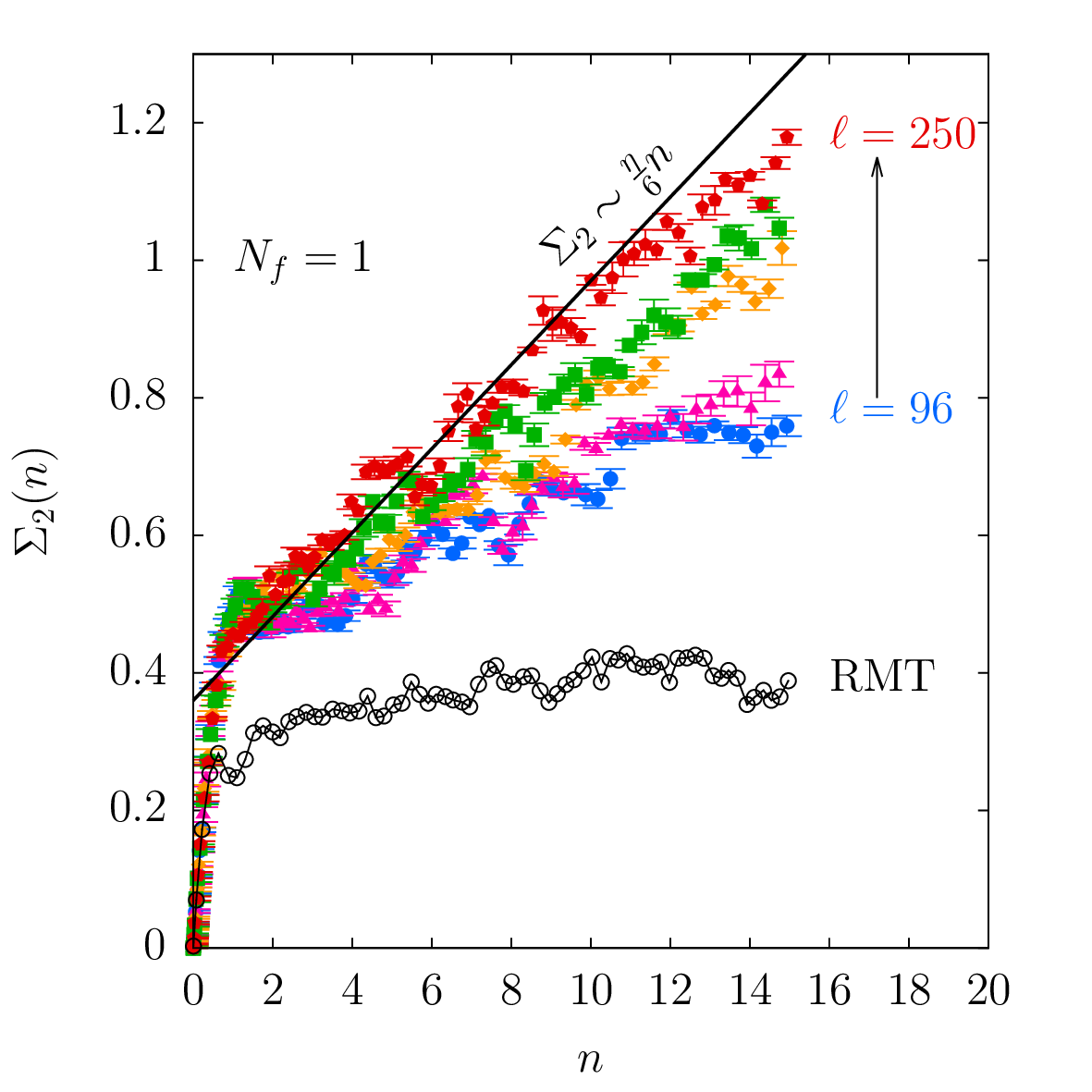}
\end{center}
\caption{(Left) The $\ell$-scaling of the inverse participation ratio $I_2$
for $N_f=1$.  The critical exponent of the scaling is $\eta=0.38(1)$.
(Right)
The number variance $\Sigma_2$ is shown as a function
of $n$.  A disagreement with nonchiral random matrix model (black points) is seen.
Instead, a critical linear rise is seen, whose slope approaches
$\eta/6$ shown as the black solid line. 
}
\label{fg:iprnvar}
\end{figure}

The Inverse Participation Ratio (IPR) is defined as 
\beq
\nonumber
I_2 \equiv \biggl\langle \int  \left(\psi^*_\lambda(x) \psi_\lambda(x)\right)^2 d^3x \biggr\rangle,
\eeq{iprdef}
where $\psi_\lambda$ is the normalized eigenvector corresponding
to the eigenvalue $\lambda$. In random matrix models, which are
ergodic, $I_2 \sim \ell^{-3}$. Thus, if the theory has a condensate,
the low-lying eigensystem of the Dirac operator would be described
by a random matrix model. Thus the IPR corresponding to the low-lying
eigenvalues should show a $\ell^{-3}$ scaling. This is another test
for the presence of $\Sigma$. Instead, if the theory is scale-invariant,
the finite size scaling of IPR would be $I_2\sim \ell^{-3+\eta}$,
where $\eta$ is a critical exponent. The exponent $\eta$ is related
to a quantity called number variance $\Sigma_2$ which measures
correlations between the eigenvalues. The number variance $\Sigma_2(n)$
is defined as the variance of the number of eigenvalues below a
value $\lambda$ which on the average contains $n$ eigenvalues.  In
ergodic random matrix models, $\Sigma_2(n)\sim\log(n)$. For a
critical theory, $\Sigma_2(n)\sim \left(\eta/6\right) n $, where
$\eta$ is the critical exponent from the IPR~\cite{Chalker:1996kr}.

In the left panel of \fgn{iprnvar}, we have shown the $\ell$-scaling
of IPR for $N_f=1$. For large $\ell$, the onset of scaling
is clearly seen. The scaling is $I_2 \sim \ell^{-2.62(1)}$. Firstly,
this rules out the ergodic $\ell^{-3}$ scaling. The theory has a
non-zero critical exponent $\eta=0.38(1)$. As explained above, in a
critical theory, $\eta$ should satisfy a critical relation to the slope of
number variance. In the right panel of \fgn{iprnvar}, we have shown
$\Sigma_2(n)$ as a function of $n$. Again, clearly there is a
disagreement with the expectation from the nonchiral random matrix theory
thereby ruling out condensate in another way. We see a
linear rise in $\Sigma_2(n)$ indicating a critical behavior. As
$\ell$ is increased, the slope of the linear rise seems to approach
$\eta/6$, as shown by the black line in the figure. Thus, both the
IPR and $\Sigma_2$ show critical behavior, and also they satisfy
the critical relation between the two.

\section{Conclusions}
In this talk, we presented convincing numerical evidences for the
absence of a bilinear condensate for all $N_f$. Instead, we found
evidences for QED$_3$ to be scale-invariant, and we estimated the mass
anomalous dimension at the infra-red fixed point at various $N_f$. In
another work~\cite{Karthik:2016bmf}, we established the presence of a
condensate in the 't Hooft limit using the same methods we described here.
This suggests an interesting phase diagram in the $(N_f,N_c)$ plane whose
one side is conformal while the other side has a mass-gap, providing a
powerful system to understand the generation of mass in QFTs. We aim to
present results on this in a future Lattice meeting.

The authors acknowledge partial support by the NSF under grant number
PHY-1205396 and PHY-1515446.

\bibliographystyle{JHEP}
\bibliography{biblio}

\end{document}